\documentclass[10pt,conference]{IEEEtran}
\usepackage{algorithmic}
\usepackage{graphicx}
\usepackage{textcomp}
\usepackage{xcolor}
\usepackage{multirow}
\usepackage{svg}
\usepackage{url}
\usepackage{threeparttable}
\usepackage{soul}
\usepackage{booktabs}
\usepackage{comment}
\usepackage{balance}
\usepackage[shortlabels]{enumitem}

\usepackage{tikz}

\newcommand\submittedtext{%
  \footnotesize This work has been submitted to and accepted by the IEEE for possible publication. Copyright may be transferred without notice, after which this version may no longer be accessible.}

\newcommand\submittednotice{%
\begin{tikzpicture}[remember picture,overlay]
\node[anchor=south,yshift=10pt] at (current page.south) {\fbox{\parbox{\dimexpr0.65\textwidth-\fboxsep-\fboxrule\relax}{\submittedtext}}};
\end{tikzpicture}%
}

\title{Language Models to Support Multi-Label Classification of Industrial Data}

\author{\IEEEauthorblockN{1\textsuperscript{st} Waleed Abdeen\IEEEauthorrefmark{1},
2\textsuperscript{nd} Michael Unterkalmsteiner\IEEEauthorrefmark{1}, 3\textsuperscript{rd} Krzysztof Wnuk\IEEEauthorrefmark{1}, \\
4\textsuperscript{th} Alessio Ferrari\IEEEauthorrefmark{2}, and
5\textsuperscript{th} Panagiota Chatzipetrou\IEEEauthorrefmark{3}}
\IEEEauthorblockA{\IEEEauthorrefmark{1}\textit{Blekinge Institute of Technology}, 
Karlskrona, Sweden \\ 
Email: waleed.abdeen@bth.se,
michael.unterkalmsteiner@bth.se,
krzysztof.wnuk@bth.se \\
\IEEEauthorrefmark{2} University College Dublin, Dublin, Ireland and CNR - ISTI, Pisa, Italy. 
Email:
alessio.ferrari@ucd.ie \\
\IEEEauthorrefmark{3}Örebro University, Sweden. Email: panagiota.chatzipetrou@oru.se
}}

\begin{document}

\maketitle

\submittednotice

\begin{abstract}
\emph{Background:} Multi-label requirements classification is an inherently challenging task, especially when dealing with numerous classes at varying levels of abstraction. The task becomes even more difficult when a limited number of requirements is available to train a supervised classifier.  Zero-shot learning does not require training data and can potentially address this problem. \emph{Objective:} This paper investigates the performance of zero-shot classifiers on a multi-label industrial dataset. The study focuses on classifying requirements according to a hierarchical taxonomy designed to support requirements tracing. \emph{Method:} We compare multiple variants of zero-shot classifiers using different embeddings, including 9 language models (LMs) with a reduced number of parameters (up to 3B), e.g., BERT, and 5 large LMs (LLMs) with a large number of parameters (up to 70B), e.g., Llama. Our ground truth includes 377 requirements and 1968 labels from 6 output spaces. For the evaluation, we adopt traditional metrics, i.e., precision, recall, $F_1$, and $F_\beta$, as well as a novel label distance metric $D_n$. This aims to better capture the classification's hierarchical nature and to provide a more nuanced evaluation of how far the results are from the ground truth. \emph{Results:} 1) The top-performing model on 5 out of 6 output spaces is T5-xl, with maximum  $F_\beta = 0.78$ and $D_n = 0.04$, while BERT base outperformed the other models in one case, with maximum $F_\beta = 0.83$ and $D_n = 0.04$. 2) LMs with smaller parameter size produce the best classification results compared to LLMs. Thus, addressing the problem in practice is feasible as limited computing power is needed. 3) The model architecture (autoencoding, autoregression, and sentence-to-sentence) significantly affects the classifier's performance. \emph{Contribution:} We conclude that using zero-shot learning for multi-label requirements classification offers promising results. We also present a novel metric that can be used to select the top-performing model for this problem.
\end{abstract}

\maketitle

\section{Introduction}
Requirements classification is one of the most common tasks in requirements engineering (RE) research~\cite{zhao2021natural}, with several contributions in the literature (e.g.,~\cite{kurtanovic_automatically_2017,alhoshan_zero-shot_2023,hey_norbert_2020}). It supports categorization for retrieval and reuse~\cite{cybulski_requirements_2000,franch_constructing_2013}, and allocation of requirements to different development teams~\cite{bashir_requirements_allocate_2023}. This paper considers the task of requirements classification intending to ensure \textit{traceability} within requirements and between requirements and other artifacts (e.g., design). This task originates from the needs of our industrial partner, which is interested in automating requirements classification to speed up the manual tracing activities. 

Requirements classification is particularly challenging when dealing with large taxonomies of possible classes, i.e., hierarchical structures of class labels and definitions thereof, and when the problem takes the form of a multi-label classification task. In RE, multi-label classification is the task of classifying (labeling) a requirement using $0\dots n$ classes from a set of classes (more than two), e.g., classifying requirements based on topics~\cite{ott_automatic_2013}. 
Unterkalmsteiner et al.~\cite{unterkalmsteiner_tt-recs_2020} call \textit{taxonomic-trace links} the problem of multi-label classification with respect to a taxonomy, with the purpose of requirements tracing. 
The problem of \textit{taxonomic-trace links} becomes particularly difficult when labeled data are scarce, as it is often the case for RE tasks~\cite{ferrari2017natural}. Supervised machine learning (ML) approaches, such as those proposed by Kurtanovic et al.~\cite{kurtanovic_automatically_2017} and Hey et al.~\cite{hey_norbert_2020} require a sufficient amount of training data for each class. Zero-shot learning is a technique that allows classification without a training phase, promising to address the data scarcity issue~\cite{alhoshan_zero-shot_2023}.

\begin{figure*}[tb]
    \centering
    \includegraphics[width=0.8\textwidth]{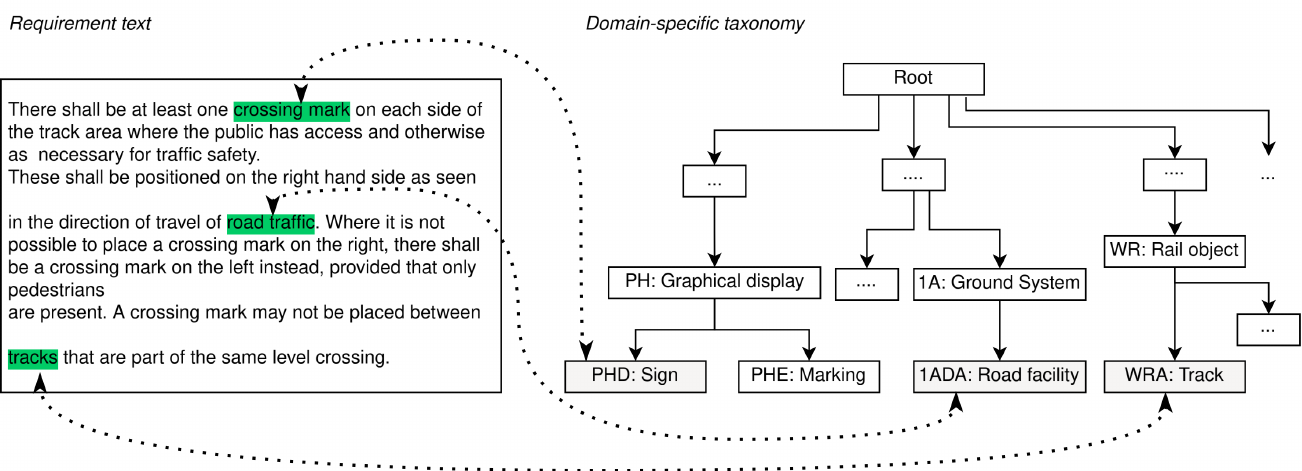}
    \caption{Example of a multi-label requirement classification task with a hierarchical taxonomy.}
    \label{fig:classification example}
\end{figure*}

Figure~\ref{fig:classification example} depicts an example of multi-label requirement classification for a taxonomic-trace links problem that was presented in our previous study~\cite{abdeen_multi_2024} in the domain of transportation software.  In this work, domain experts annotated the requirement using classes from a large domain-specific taxonomy. Three classifications were made to the requirement, \emph{sign}, \emph{road facility}, and \emph{track}, based on the terms mentioned in the requirement text: \emph{crossing mark}, \emph{road traffic}, and \emph{tracks}. Eventually, the requirement text as a whole received three classifications (labels), leading to the formulation as a multi-label classification problem. 

Language models (LM)~\cite{vaswani_attention_2017} are statistical models of natural language, typically based on neural networks. Large LMs (LLMs) are LMs with large sets of parameters, in the order of billions. They are pre-trained on large corpora to process, understand, and generate text. 
The knowledge that these pre-trained models have can be transferred to downstream natural language processing (NLP) tasks without training (zero-shot) or by having a relatively small labeled dataset to fine-tune the models~\cite{devlin_bert_2019}. 

Our main objective is determining the best-performing LM or LLM for a zero-shot classifier to label a set of requirements from a large set of possible hierarchical classes. We conducted an experiment to compare 14 state-of-the-art and top-ranked LMs and LLMs to generate text embeddings for the classifier. We evaluate the classifier's performance using a dataset of 377 requirements labeled with 1968 classes~\cite{abdeen_multi_2024}. To measure the performance of the classifier, we use traditional IR metrics (precision, recall, and $F_{1}$-score), context-dependent metric $F_{\beta}$, and a novel \emph{label distance metric}, which takes into account the hierarchical nature of the classification problem.

The paper is organized as follows. In Section~\ref{sec:background and rw}, we present background and related work. We explain the classification problem and introduce the zero-shot classifier based on LMs in Section~\ref{sec:classification}. In Section~\ref{sec:experiment}, we present the experimental setup, while in Section~\ref{sec:results}, we present the results and discuss them in Section~\ref{sec:discussion}. Section ~\ref{sec:conclusion} concludes the paper.

\section{Background and Related Work}\label{sec:background and rw}

In this section we present background information for LMs, and LLMs, and related work in multi-label requirements classification and zero-shot learning.

\subsection{Language Models}

Language models (LMs)~\cite{vaswani_attention_2017} are statistical models trained on large corpora to process, understand, and generate natural language. They support various NLP tasks, such as text understanding, topic modeling, classification, named-entity recognition, and text generation. The Transformer model, introduced by Vaswani et al. in 2017 ~\cite{vaswani_attention_2017}  uses an encoder to convert input text into continuous vectors and a decoder to generate output. The Transformer is the foundational architecture of many modern LMs. It employs self-attention to assign contextual weights to words, enhancing comprehension. BERT, a bidirectional Transformer, combines left and right context in pre-training, achieving state-of-the-art performance in NLP tasks \cite{kici_bert-based_2021,bashir_requirements_allocate_2023,alhoshan_zero-shot_2023,devlin_bert_2019}.

LLMs are LMs pre-trained on much larger datasets and optimized for text generation, allowing them to predict the next word based on previous text ~\cite{brown_language_2020,touvron_llama_2023,jiang_mixtral_2024}. Similar to LMs, LLMs typically use the Transformer architecture. OpenAI's GPT-3~\cite{brown_language_2020}, with 175 billion parameters, was introduced in 2022, and achieved performance comparable to state-of-the-art fine-tuned systems. More recently, Meta's Llama 2 ~\cite{touvron_llama_2023}, ranging from 7 to 70 billion parameters, was fine-tuned with reinforcement learning via human feedback, yielding higher performance than GPT-3. Additionally, models like Mixtral ~\cite{jiang_mixtral_2024} combine multiple LLMs (eight 7B models) into a 47B-parameter model, outperforming larger models like Llama2 70B, suggesting that parameter count alone doesn’t determine performance.

LMs and LLMs generate word embeddings—numerical representations of text in n-dimensional space. Although word2vec first introduced this concept~\cite{jurafsky_vector_2024}, LMs and LLMs produce more advanced embeddings by using self-attention to consider previous token context~\cite{vaswani_attention_2017}. 
These models improve the performance compared to word2vec models in several tasks~\cite{schopf_evaluating_2022}.

These models can also be fine-tuned on specific tasks, allowing them to achieve higher accuracy by adapting to particular datasets. Alternatively, they can transfer learned knowledge to other tasks, enabling zero-shot classification, where pre-trained models are used to classify new types of data without additional task-specific training.

\subsection{Related Work}
In the following, we summarise related works in the area of requirements classification, with a focus on the use of LMs and LLMs, zero-shot solutions, and multi-label classification. 

Kici et al.~\cite{kici_bert-based_2021} used the transfer learning approach based on LMs to classify requirements specifications on three dimensions (type, priority, and severity), containing between four and 21 classes. They fine-tuned the BERT model using labeled software requirements specifications. The LM-based approach produced a higher $F_1$-score than traditional ML approaches. Also using BERT, Hey et al.~\cite{hey_norbert_2020}, used transfer learning across different non-functional requirements classification tasks, achieving state-of-the-art performance in the RE field compared to previous work.  

Bashir et al.~\cite{bashir_requirements_allocate_2023} experimented with different classification pipelines based on LMs to assign requirements to specific teams. They fine-tuned BERT-based LMs on 15 classes, each representing a team on an industrial dataset. The BERT-based classifier was also superior to traditional supervised ML approaches (e.g., SVM), with an average of 68\% $F_1$-score. Moreover, 65\% of the domain vocabulary in their dataset was not found in the dataset used to train the BERT model. The authors suggest that domain-specific LMs could lead to better performance.

Alhoshan et al.~\cite{alhoshan_zero-shot_2023} is the first work in RE applying zero-shot learning to classify requirements without labeled training data, using pre-trained LMs to generate embeddings for requirements and classes. They compared LMs across three tasks: NFR vs. functional classification, NFR classification, and security requirements identification. The zero-shot classifier achieved up to an 89\% $F_{1}$-score on top NFR classes. Overall, general-purpose LMs outperformed domain-specific LMs optimized for requirements tasks.

Among the few works on multi-label classification, Ott~\cite{ott_automatic_2013} classified automobile requirements into 141 topics, with some requirements covering multiple topics, hence multi-label. They trained an ML model on a large labeled dataset. The trained ML model was able to achieve up to 83\% recall and 66\% precision. However, the transfer of the model to another dataset caused recall and precision to drop to 40\% and 50\%, respectively.

In a previous study~\cite{abdeen_multi_2024} we explored multi-label requirements classification according to a hierarchical taxonomy, examining the performance of a zero-shot classifier based on Explicit Semantic Analysis (ESA)~\cite{gabrilovich_computing_2007}. The classifier performance improved compared to a word2vec-based classifier that was developed earlier~\cite{unterkalmsteiner_tt-recs_2020}. However, the absolute performance was limited.
Furthermore, the study did not leverage the potential of LMs and LLMs for zero-shot classification. 

These previous works show the interest of the RE field in requirements classification, some preliminary uses of LMs for this task, and also some limited contributions to the multi-label classification problem. However, none of the studies have used LMs and LLMs for the task of multi-label requirements classification with large taxonomies. This study aims to fill this gap with a systematic comparison of models for this task. We apply zero-shot classification to address the data scarcity problem typical of the RE field, and we use an industrial dataset, which strengthens the practical relevance of our contribution. 

\section{Multi-Label Requirements Classification}\label{sec:classification}
Large-scale multi-label requirements classification concerns assigning $0 \dots n$ labels from a large output space to a requirement artifact. In this section, we present the classification task, and the classifier used to accomplish the task.

\subsection{The Classification Task}

We classify natural language requirements using a domain-specific hierarchical taxonomy. The input includes the requirement text and additional context, like document and section titles. The output is a set of domain-specific concepts from a taxonomy, similar to the one in Figure~\ref{fig:classification example}. The nodes in the taxonomy are our \textit{labels} (i.e., classes) and are all associated with textual descriptions. 

\subsection{Zero-Shot Requirements Classifier}\label{sec:classifier}
\begin{figure*}[htb]
    \centering
    \includegraphics[width=0.9\textwidth]{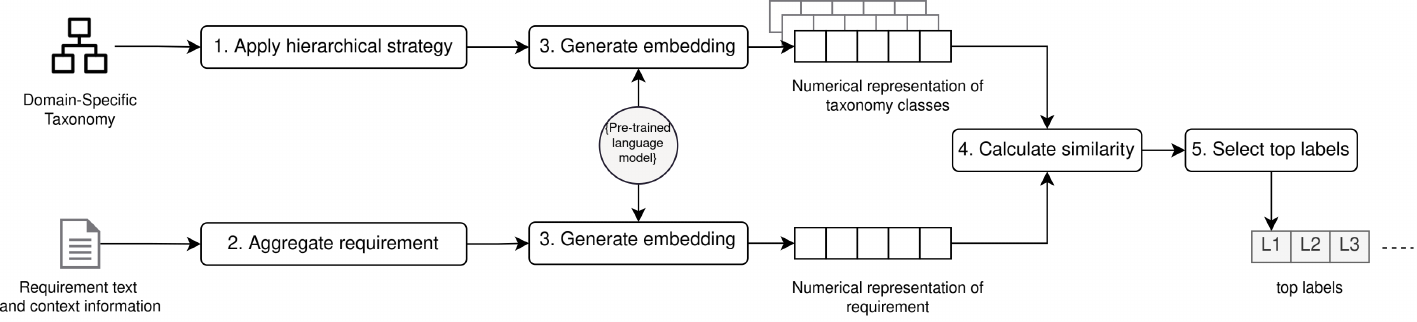}
    \caption{Zero-Shot Classifier}
    \label{fig:zsc}
\end{figure*}

In this study, we use pre-trained models to generate embeddings for a zero-shot requirements classifier. The pipeline components (see Figure~\ref{fig:zsc}) of the classifier are explained in the following.

\begin{enumerate}[1.]
    \item \textbf{Apply hierarchical strategy:} A pre-processing step is performed on the taxonomy by aggregating the descriptions of the child nodes with those of the parent node. This strategy, which we call ``hierarchical strategy'' has shown to be effective in increasing the performance of text classification tasks with a hierarchical output space ~\cite{abdeen_multi_2024}. In practice, this allows us to include additional semantic information on child nodes, which can be exploited by the classifier to better understand the meaning of the class represented by the node. 
    \item \textbf{Aggregate requirement:} The requirement text is aggregated with context information, i.e., document and section titles where the requirement is specified. The aggregated text is used to generate the embedding for the requirement. As we saw from examining requirements samples, the context information is necessary to understand many of the requirements. Thus, we use this information as input to the model to generate more representative embeddings.
    \item \textbf{Generate embedding:} An embedding is generated for both the aggregated requirement text and the aggregated nodes description for each class in the taxonomy, using LMs or LLMs.
    \item \textbf{Calculate similarity:} A cosine text similarity is calculated between the requirement text embeddings and each class embedding, to show how close these embeddings are to each other. 
    \item \textbf{Top labels selection:} The classes with the highest similarity scores are selected as labels for the requirement. The number of labels is specified by the user.
\end{enumerate}

\section{Experiment Setup}\label{sec:experiment}

We design the experiment following the experimentation guidelines by Wohlin et al.~\cite{wohlin_experimentation_2012} and report it using the experiment reporting framework proposed by Jedlitschka et al.~\cite{jedlitschka_reporting_2008}. A replication package including the classifier, experiment code, and annotated requirements is available in our replication package\footnote{https://doi.org/10.6084/m9.figshare.25990825.v1}.

\subsection{Goals Questions and Metrics}
The experiment aims to explore the effect and performance of using LMs and LLMs in a zero-shot classifier for a multi-label requirements classification task. In alignment with this aim, we construct a Goal-Question-Metric (GQM) matrix as follows:

\begin{enumerate}[G1:]
    \item \emph{Find the model that produces requirements text embedding that leads to the best classifier’s performance.}

    \item \emph{Identify the model characteristics that impact the classifier’s performance.}

\end{enumerate}

In alignment with these goals, we wrote the following three research questions mapped one-to-one to the goals.

\begin{enumerate}[RQ1:]
    \item  \emph{To what extent does the zero-shot classifier’s performance differ when using embedding created by different models?}
    
    Motivation: A zero-shot classifier benefits from pre-trained models to generate text embedding. Due to the increasing number of pre-trained models that perform well on different software-related tasks, we ask this question to find the best-performing LM or LLM. 
    
    \item \emph{What model’s characteristics are associated with the classifier performance?}

    Motivation: Using LMs and LLMs in a zero-shot classifier could lead to different results. Exploring the model’s characteristics (embedding size, model size, and model type) that have an impact on the classifier’s performance can facilitate the choice of the model for similar classification problems.

\end{enumerate}

We use traditional metrics \emph{precision, recall, $F_{1}$-score and $F_{\beta}$} 
(Section~\ref{sec:weighted f-score}) as well a novel \emph{label distance metric} (Section~\ref{sec:distance}), to measure the classifier’s performance, thus answering RQ1 and RQ2.

\subsection{Materials}
The experiment’s materials consist of a dataset, available in our replication package, and taxonomies from our previous study~\cite{abdeen_multi_2024}.

\subsubsection{Dataset} The whole dataset that our industrial partner want to classify contains 22000 requirements. For the purpose of our study, we sampled and used a subset of 377 requirements, which we annotated with 1968 labels using two taxonomies. This sample size was calculated with a 95\% confidence interval and a ±5\% margin of error, ensuring that the sample estimates are accurate and represent of the whole population. The sampled requirements consist of 340 general regulatory requirements and 37 project-specific requirements. We used the requirement text and context information (document and section titles) as input to the classifier. The length of a the classified text ranges between 45 to 813 characters. We used the dataset as a test set to measure the classifier's performance.

\subsubsection{Taxonomies} We classify the requirements using two domain-specific taxonomies (SB11 and CoClass), each consisting of three dimensions describing orthogonal aspects of the problem domain. These taxonomies' nodes (classes) are domain concepts on different abstraction levels, and each dimension consists of 250 to 1183 nodes. In our experiment, we treat each
dimension as a separate output space and, consequently, have six output spaces in total. We list the characteristics of the output spaces in Table~\ref{tab:os_ch}. 

\begin{table}[h]
    \caption{Output spaces’s characteristics}\label{tab:os_ch}
    \centering
    \begin{threeparttable}
    \footnotesize
    \resizebox{1\columnwidth}{!}{
    \begin{tabular}{lllllll}
            \toprule
             OS~\tnote{a} & Desc. length \tnote{b} & Depth & Categories & Leaf nodes & Total nodes \\
            \midrule
             A & 27 & 5 & 50 & 206 & 256 \\ 
             B & 28 & 6 & 299 & 884 & 1183 \\
             G & 96 & 3 & 199 & 665 & 864 \\
             K & 92 & 3 & 61 & 251 & 312  \\
             L & 40 & 1 & 0 & 635 & 635 \\ 
             T & 79 & 4 & 80 & 170 & 250 \\ \bottomrule
        \end{tabular}
    }
    \begin{tablenotes}
        \item[a] OS: output space
        \item[b] Description length: mean characters count of nodes description
    \end{tablenotes}
    \end{threeparttable}
\end{table}

\subsection{Variables and Hypotheses}

We have one independent variable, the model used to generate embedding, with 14 treatments. We experiment with state-of-the-art models that show high performance on other RE and non-RE tasks. We complement with models that were on the sentence-transformer’s leaderboard\footnote{\url{https://www.sbert.net/docs/pretrained_models.html}}. We list the models with their specification in Table~\ref{tab:models specs}. These models are based on the transformer’s architecture, in which a model consists of an encoder and a decoder. We recognize three main model types based on their architecture:

\paragraph{Autoencoding}
Consists of only the encoder part of the original transformer model. These models are usually used for sentence or token classification. We experimented with six models in this category: MiniLM~\cite{wang_minilm_2020} and DistillRoBERTa~\cite{sanh_distilbert_2020} are distilled versions of BERT and RoBERTa models, respectively. They have a lower number of parameters and require fewer resources to run than the original models. MiniLM was the top-performing model on NFR identification and classification tasks~\cite{alhoshan_zero-shot_2023}.
MPNet~\cite{song_mpnet_2020} is a pre-trained language model that leverages the dependency among the predicted tokens and accounts for more information when predicting masked tokens. MPNet has outperformed BERT and RoBERTa on language understanding tasks~\cite{song_mpnet_2020}. Also, it is the top-performing model on the sentence embedding and text classification tasks according to the sentence-transformer's leaderboard.
RoBERTa models (base and large)~\cite{liu_roberta_2019}, are robustly optimized BERT models that are trained on a larger dataset than the original BERT model, and outperform BERT on the language understanding task.

\paragraph{Autoregression}
Consists of the decoder part only of the original transformer model. The common use for these models is text generation.
Llama2~\cite{touvron_llama_2023} is an LLM with up to 70B parameters, pre-trained on 2 trillion tokens. Llama2 is the largest among the experimented models in terms of learned parameters and the dataset size on which it was pre-trained.
Mistral-7B~\cite{jiang_mistral_2023} is an LLM that adapts different techniques to keep the high performance of LLMs while making them more efficient to run. Mistral-7B has matched or outperformed Llama2-13B ($\approx 2x$  its size) on knowledge, reasoning, and comprehension tasks~\cite{jiang_mistral_2023}.
Mixtral-8x7B~\cite{jiang_mixtral_2024} is a sparse mixture of expert language models based on Mistral-7B architecture. The model combines the knowledge of multiple expert models to provide the best results. Mixtral outperformed Llama2-70B in mathematics, code generation, and multilingual benchmarks.

\paragraph{Sequence-to-sequence}
Models containing both an encoder and a decoder are designed for translation, summarization, and question-answering tasks.
T5~\cite{ni_sentence-t5_2021} was the first model to use both the encoder and decoder to create sentence embedding. The model outperformed other models that were optimized for sentence embedding, mainly the Sentence-BERT (SBERT) model~\cite{reimers-2019-sentence-bert}.

We loaded the autoencoding and sequence-to-sequence models from a checkpoint where the models are fine-tuned for sentence embedding and are referred to as sentence transformers. Previous studies~\cite{reimers-2019-sentence-bert} showed that sentence-based classifiers perform better and more efficiently than token-based classifiers on text classification tasks.

\begin{table*}[bth]
    \centering
    \caption{Detailed specifications of the language models used in the experiment (independent variable treatments)}
    \resizebox{1\textwidth}{!}{
    \footnotesize
    \begin{tabular}[\textwidth]{lllllll}
        \toprule
         Model Type & Model & Checkpoint & Model size (params) & Context & Embedding size & KB (dataset) \\ 
         \midrule
         \multirow{6}{*}{Autoencoding}  & BERT (base) & all-MiniLM-L12-v2~\cite{wang_minilm_2020} &  66M & 512 & 384 & $\succ$ 1B sentences pairs \\ 
         & DistillRoBERTa (base) & all-distilroberta-v1 & 82.8M & 128 & 768 & $\succ$ 1B senetence pairs \\
         & MPNet (base)~\cite{song_mpnet_2020} & all-mpnet-base-v2 & 133M & 512 & 768 & $\succ$ 1B sentences pairs \\
         & MPNet (base) & multi-qa-mpnet-base-dot-v1 & 133M & 512 & 768 & 215M question, answer pairs\\ 
         & RoBERTa (base) & msmarco-roberta-base-v3 & 125M & 510 & 768 & $\succ$ (160GB) 1B sentence pairs \\
         & RoBERTa (large) & all-roberta-large-v1 & 355M & 128 & 1024 & $\succ$ 1B sentence pairs \\ 
         \midrule
         \multirow{5}{*}{Autoregression} 
         & Llama2-7B & Llama2-7B & 7B & 4096 & 4096 &  2 trillion tokens \\
         & Llama2-13B & Llama2-13B & 13B & 4096 & 4096 &  2 trillion tokens \\
         & Llama2-70B & Llama2-70B-Q4 & 70B & 4096 & 4096 & 2 trillion tokens \\
         & Mistral-7B & Mistral-7B & 7B & 4096 & 4096 & N/A\\
         & Mixtral-8x7B & Mixtral-8x7B & 47B & 4096 & 4096 & N/A\\
         \midrule
         \multirow{3}{*}{Sequence-to-sequence}
         & T5-base & sentence-t5-base & 220M & no limit & 768 & 2.6M instances\\
         & T5-large & sentence-t5-large & 770M & no limit & 768 & 2.6M instances\\
         & T5-xl & sentence-t5-xl & 3B & no limit & 768 & 2.6M instances\\
         
        \bottomrule
    \end{tabular}
    }
    \label{tab:models specs}
\end{table*}

We make the following hypotheses for our experiment and align them with the research questions in Table~\ref{tab:ex_alignment}.
\begin{itemize}
    \item $H_{A0}$ The model used to generate embedding does not significantly impact the classifier’s performance.
    \item $H_{A1}$ The model used to generate embedding significantly impacts the classifier’s performance.
    \item $H_{B0}$ The embedding size generated by a model does not significantly impact the classifier’s performance.
    \item $H_{B1}$ The embedding size generated by a model significantly impacts the classifier’s performance.
    \item $H_{C0}$ The model size (number of parameters learned) has no correlation with the classifier’s performance.
    \item $H_{C1}$ The model size (number of parameters learned) positively correlates with the classifier’s performance.
    \item $H_{D0}$ The model type (autoencoding, autoregression, sequence-to-sequence) does not significantly impact the classifier’s performance.
    \item $H_{D1}$ The model type (autoencoding, autoregression, sequence-to-sequence) significantly impacts the classifier’s performance.
\end{itemize}

\begin{table}[htb]
    \centering
    \caption{Experiment design alignment}
    \resizebox{1\columnwidth}{!}{
        \begin{tabular}{cccccc}
        \toprule
             RQ & Hypothesis & Independent var.& Dependent var. & Metric & Analysis \\
        \midrule
             RQ1 & $H_{A}$ & Embedding size & Performance & $F_{1}$, $F_\beta$, $D_a$ & Comparative analysis\\
             RQ2 & $H_{B}$ & Embedding size & Performance & $F_{1}$, $F_\beta$, $D_a$ & Kruskal-Wallis H \\
             RQ2 & $H_{C}$ & Model size & Performance & $F_{1}$, $F_\beta$, $D_a$ & Spearman's coefficient\\
             RQ2 & $H_{D}$ & Model type & Performance & $F_{1}$, $F_\beta$, $D_a$ & Kruskal-Wallis H \\
        \bottomrule
        \end{tabular}
    }
    \begin{tablenotes}
        \footnotesize
        \item Var. : variable
    \end{tablenotes}
    \label{tab:ex_alignment}
\end{table}

\subsection{Experiment Design}
We designed our experiment as one-factor, the model, with 14 treatments. We evaluated each treatment using six subsets of requirements, each annotated with one output space $OS_{A}$, $OS_{B}$, $OS_{G}$, $OS_{K}$, $OS_{L}$ and $OS_{T}$. In total, we run 14 * 6 = 84 experiment instances.

\subsection{$F_{\beta}$-score}
\label{sec:weighted f-score}

Similar to previous studies, we use the standard IR metrics, i.e., precision, recall, and $F_1$-score, to evaluate the performance of the classifier. However, using these metrics may not be the best way to validate the classifier for the requirements tracing task, as this can be considered a ``hairy'' RE task, which requires more refined metrics. Berry~\cite{berry_empirical_2021} defines a hairy RE task as follows:

\begin{quote}
 \textit{``A hairy requirements engineering (RE) task involving natural language (NL) documents: 1) is a non-algorithmic task that requires NL understanding~\cite{ryan_role_1993} and 2) is not difficult for humans to do on a small scale but 3) is unmanageable when it is done to the documents that accompany the development of a large computer-based system (CBS)~\cite{northrop2006ultra}."}
\end{quote}

We argue that multi-label requirements classification of natural language requirements can be considered a hairy RE task due to the following reasoning. Requirements classification can't be logically guaranteed, but rather, a process that involves an understanding of the requirement content is required to find the correct classes for a requirement from a large set of classes. Engineers can perform the task on a small scale. However, it is still a resource-intensive, and even for domain experts, challenging task~\cite{unterkalmsteiner_tt-recs_2020}. The task does not scale easily, i.e., for hundreds or thousands of requirements, as multiple labels need to be created and maintained for each requirement. On such a hairy RE task, traditional information retrieval (IR) metrics (precision, recall, and $F_{1}$-score) may not be the best option to evaluate the classifier performance, but rather a context-dependent evaluation shall be considered~\cite{berry_panel_2017}. In such an evaluation, researchers should determine the most important aspect, precision or recall, for the RE task for which the tool (the zero-shot classifier) would be used~\cite{berry_panel_2017}. Then, a weighted F-score is calculated, based on the important aspect of the hairy task, rather than calculating an $F_{1}$-score, where both precision and recall are weighted equally. The weighted F-score ($F_{\beta}$) is calculated using the formula:

\begin{equation}
    F_{\beta} = \frac{(1+\beta^2)  \times Precision \times Recall} {(\beta^2 \times Precision) + Recall}
\end{equation}

where $\beta$ is a weighting coefficient that, with an increase, emphasizes recall and, with a decrease, emphasizes precision. Berry~\cite{berry_empirical_2021} presented multiple techniques to estimate $\beta$, depending on the available data from building the ground truth. One way of calculating $\beta$, to achieve task \emph{T} on a set of documents \emph{D}, is by using the formula:

\begin{equation}
    \beta_{D,T} = \frac{l}{\lambda_{D,T}}
\end{equation}

where \emph{l} is the number of all potential answers (labels) for the ground truth (G) and $\lambda_{D,T}$ is the true answers (labels) in G. In our case, an answer is a label. Consequently, \emph{l} is the number of potential labels (from the taxonomy) for the requirements dataset. While building the ground truth, we classified the requirements using six output spaces (dimensions) from two taxonomies. Thus, we calculate the number of potential labels for each output space and present the results in Table~\ref{tab:beta}. As an example, we introduce the detailed calculation of $\beta$ for the output space $OS_{T}$.

\begin{equation}
    l = l_{head} \times l_{tail} = 81 \times 250 = 20250    
\end{equation}

where $l_{head}$ is the dataset size (G), which is the number of requirements classified using $OS_{T}$ in G, and  $l_{tail}$ is the number of potential labels, which is the number of $OS_{T}$ classes.

\begin{equation}
    \beta_{D,T} = \frac{l}{\lambda_{D,T}} = \frac{20250}{107} = 189.25 
\end{equation}

where $\lambda_{D,T}$ is the number of labels assigned to the requirement in G. So the minimum value of $\beta$ should be 189.25, which implies that recall is more important than precision. Thus, we calculate $F_{189}$ to evaluate the use of the classifier for the classification task on $OS_T$.

\begin{table}[htb]
    \centering
    \caption{$\beta_{D,T}$ calculated for each output space}
    \footnotesize
    \begin{tabular}{ccccc}
        \toprule
        OS & $l_{head}$ & $l_{tail}$ & $\lambda_{D,T}$ & $\beta$  \\ 
        \midrule
        A & 24 & 256 & 30 &  205 \\
        B & 123 & 1183 & 236 & 617 \\
        G & 99 & 864 & 175 & 489 \\
        K & 59 & 312 & 70 & 263 \\
        L & 56 & 635 & 78 & 456\\
        T & 81 & 250 & 107 & 189 \\
        \bottomrule
    \end{tabular}
    \label{tab:beta}
\end{table}

\subsection{Label Distance Metric}\label{sec:distance}

In this section, we present the \emph{label distance metric}, a more granular measure of the requirements classifier’s performance. We motivate the need for a new metric, present it, and explain how it is measured. Eisner et al.~\cite{eisner_improving_2005} have used a similar distance metric to analyze the results of a hierarchical classifier of protein function. We adapt this metric for this task.

\subsubsection{Motivation}
Traditional metrics, precision, recall, and $F_{1}$-score, are calculated based on TP, TN, FP, and FN, which are binary metrics, i.e., a predicted label can either be true positive, true negative, false positive, or false negative. These traditional metrics are well-suited if the labels have no semantic relationship to each other. However, in a hierarchical classification scenario, we can exploit the relationship between labels to create a more nuanced metric that considers the distance (for example, expressed as the number of nodes) between labels in a hierarchy. Such a metric would be more sensitive, as it provides continuous values compared to traditional metrics that encode only a binary truth value. In other words, such a metric could express the distance between predicted and true values instead of encoding just hit or miss. 
This is particularly important for hairy RE problems where a classifier may not provide the exact answer but a close enough one for an engineer to make a decision in human-in-the-loop~\cite{parasuraman2000model} automation.

\subsubsection{Proposed Metric}

\begin{figure}[htp]
    \centering
    \includegraphics{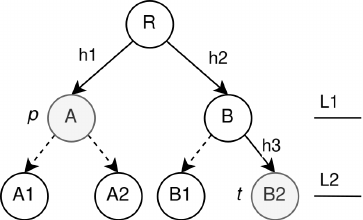}
    \caption{An example demonstrating the measurement of the label distance metric in a taxonomy between p (predicted label) and t (true label)}
    \label{fig:distance}
\end{figure}

We propose the \emph{label distance metric}, a measurement of how close the predicted label is to the true label. We use the example in Figure~\ref{fig:distance} to explain how the distance metric can be realized. 
The normalized distance metric value ranges between zero, the predicted label matches the true label, and one, the predicted label is the furthest away possible from the true label in the output space. The distance calculation is dependent on the output space used for classification.  We calculate the distance between the predicted and true label using the formula:

\begin{equation}
    D_a = f(p,t)
\end{equation}
		
where $D_a$ is the absolute distance measure, $p$ is the predicted label, $t$ is the true label, and $f$ is the function used to calculate the distance between the two labels. In Figure~\ref{fig:distance}, $p$ is node $A$ and $t$ is node $B2$. The distance between classes can be measured by counting the number of hops from one class in the hierarchy to another:

\begin{equation}\label{eq:distance function}
    f(p,t) = \sum h (p,t)
\end{equation}
		
where $h$ is a hop between the labels $p$ and $t$.
After measuring the distance, we calculate a normalized distance $D_{n}$ between 0 and 1 to ensure that the distance can be used to compare the performance of different classifier designs on different output spaces. We measure $D_{n}$ using the formula:

\begin{equation}
    D_{n} = \frac{D_a}{D_{max}}
\end{equation}
		
where $D_{max}$ is the maximum distance, i.e., the maximum number of hops to reach from one class to another in the tree. $D_{max}$ is calculated using the formula:

\begin{equation}
    D_{max} = 2 \times depth
\end{equation}

where depth is the maximum depth level of the tree, the distance is calculated per predicted label/true label pair and then averaged across all labels. 

\subsubsection{Example}

Assuming that a requirement has \emph{A} as a true label from a taxonomy (Figure~\ref{fig:distance}), and a classifier predicted \emph{B2} label for the requirement. The number of hops from \emph{A} to \emph{B2} in the taxonomy is three, and the taxonomy has a depth of two. So, we calculate the absolute distance as follows:

\begin{equation}
    D_a = f(p,t) = \sum h (p,t) = \sum h (A,B2) = 3
\end{equation}

and the normalized distance:

\begin{equation}
     D_{max} = 2 \times depth = 2 \times 2 = 4 
\end{equation}
\begin{equation}
     D_{n} = \frac{D_a}{D_{max}} = \frac{3}{4} = 0.75
\end{equation}

The normalized distance between the true label \emph{A} and the predicted label \emph{B2} is 0.75. Using the label distance metric we were able to measure how far the predicted label is from the true label, without manually analyzing each prediction.

\subsection{Analysis Procedure}\label{sec:analysis procedure}

Normality was assessed by Shapiro-Wilk's test, and none of our variables follows a normal distribution ($p<0.05$). We use Spearman's correlation to determine whether there is an association between the model size and the classifier’s performance (both measured on a continuous scale). Spearman's correlation is a nonparametric test that measures the strength and the direction of the association between two continuous variables. 

Moreover, we use Kruskal-Wallis H test. Kruskal-Wallis H test is a rank-based nonparametric test that can be used to determine if there are statistically significant differences between two or more groups of an independent variable on a continuous dependent variable. We use Kruskal-Wallis H test to understand the effect of using different model types (autoencoding, autoregression, and sequence-to-sequence) and the effect of embedding size (384, 768, 1024, 4096) on the classifier’s performance~\cite{conover1999practical,wohlin_experimentation_2012}.  

However, finding an association using Kruskal-Wallis H test (i.e., $p<0.05$) indicates that the median of at least one group is different from the median of another group and did not provide us with further details about this association (e.g., which groups are ’responsible’ for this association). Therefore, following up on our statistically significant results, we performed post hoc testing using Dunn's~\cite{dunn1964multiple} procedure with a Bonferroni adjustment. The alignment of the statistical analysis with the research questions and hypotheses can be found in Table~\ref{tab:ex_alignment}.

\subsection{Threats to Validity}

There are four main validity aspects to this study, namely construct, internal, external, and conclusion validity~\cite{jedlitschka_reporting_2008,wohlin_experimentation_2012}.

\textit{Construct validity.} A threat to the construct validity of the study is the wrong design of the experiment. We could have missed a confounding variables, or chose an in appropriate statistical test. We tried to mitigate this threat by following best practices for designing~\cite{wohlin_experimentation_2012} and reporting experiments~\cite{jedlitschka_reporting_2008}.

\textit{Internal Validity.} A threat to the internal validity is the potential errors when performing the experiment. To mitigate this treat, we publish a replication package containing the source code of the classifier and the experiment. 

\textit{External validity.} The dataset is from one domain. Future studies are needed to generalize the results in other domains. 

\textit{Conclusion validity.} Our study's conclusion is dependent on the analysis of the results using the Kruskal–Wallis test and scatter plots. Kruskal–Wallis is a non-parametric test, a good fit when there are no assumptions about the data distribution; however, it is less accurate than parametric tests. Moreover, analyzing the results using scatter plots is subjective.
\subsection{Task}
We run all experiment instances on a local machine. We used the sentence transformers and langchain libraries to generate embeddings. Due to resource limitations we used the quantized version q4 from Llama2 70B, and Mistral 7B and Mixtral 8X7B models.

The zero-shot classifier produces k labels per requirement. We experimented with different values of k but report, due to space constraints, only the results at k = 15. The higher the k, the better the recall and the worse the precision of the classifiers, all other things being equal. In practice, one can choose k to trade off the effort of reviewing suggested labels with the risk of missing relevant labels. With output spaces comprising 250 to 1183 classes, setting k to 15 means a reduction between 94\% and 99\% in the number of options presented to the requirements engineer as possible classes.

\section{Results}\label{sec:results}

Table~\ref{tab:results_prf} depicts the results of the classifier performance using precision, recall, $F_{1}$, $F_{189}$ and $D_n$. The bold values represent the top results between all instances, while the underlined values are the top results on a specific dataset (A,B,G,K,L,T). We did not calculate $F_{\beta}$ according to $\beta$ of each output space as presented in Table~\ref{tab:beta}, but rather calculated $F_{189}$, which is according to the smallest output space. This is because the value of $F_{\beta}$ approximates the value of recall when $\beta \approx 200$ or above, which was also observed by Berry~\cite{berry_empirical_2021}. 

\begin{table*}[bthp]
\caption{Classifier’s performance results (precision, recall, f1) @ K=15}
\begin{center}

\begingroup
\setlength{\tabcolsep}{2pt} 
\renewcommand{\arraystretch}{1.2} 

\resizebox{1\textwidth}{!}{
\begin{tabular}{l|rrrrr@{\hskip 0.4cm}rrrrr@{\hskip 0.4cm}rrrrr@{\hskip 0.4cm}rrrrr@{\hskip 0.4cm}rrrrr@{\hskip 0.4cm}rrrrr}

\toprule
    ~ & A & ~ & ~ & ~ & ~ & B & ~ & ~ & ~ & ~ & G & ~ & ~ & ~ & ~ & K & ~ & ~ & ~ & ~ & L & ~ & ~ & ~ & ~ & T & ~ & ~ & ~ & ~ \\ 

    Model & P & R & $F_1$ & $F_{189}$ & $D_n$ & P & R & $F_1$ & $F_{189}$ & $D_n$ & P & R & $F_1$ & $F_{189}$ & $D_n$ & P & R & $F_1$ & $F_{189}$ & $D_n$ & P & R & $F_1$ & $F_{189}$ & $D_n$ & P & R & $F_1$ & $F_{189}$ & $D_n$ \\ 
    \midrule
    BERT-base (MiniLM) & 
    \underline{\textbf{.07}} & \underline{\textbf{.83}} & \underline{\textbf{.12}} & \underline{\textbf{.83}} & 
    \underline{\textbf{.04}} &
    .05 & .38 & .09 & .38 & .14 &
    .05 & .48 & .09 & .48 & .20 &
    .04 & .59 & .08 & .59 & .18 &
    \underline{.05} & .50 & .08 & .50 & .45 &
    \underline{\textbf{.07}} & .75 & \underline{\textbf{.12}}& \underline{.75} & \underline{.08}
    \\
    
    DistillRoBERTa &
    .05 & .64 & .09 & .64 & .08 &
    .05 & .40 & .09 & .40 & .13 &
    .05 & .45 & .09 & .45 & .18 &
    .05 & .67 & .09 & .67 & .11 &
    \underline{.05} & .52 & \underline{.09} & .52 & .45 &
    .06 & .62 & .10 & .62 & .12
    \\
    MPNet &
    .06 & .74 & .11 & .74 & .05 &
    .04 & .35 & .08 & .35 & .13 &
    .05 & .47 & .09 & .47 & .19 &
    .05 & .63 & .09 & .63 & .14 &
    .04 & .47 & .08 & .47 & .50 &
    .06 & .69 & .11 & .69 & .12
    \\
    Multi-MPNet &
    .05 & .67 & .10 & .67 & .08 &
    .05 & .40 & .09 & .40 & .14 &
    .05 & .44 & .09 & .43 & .21 &
    .05 & .64 & .09 & .64 & .14 &
    \underline{.05} & .51 & .08 & .51 & .47 &
    .05 & .57 & .09 & .57 & .17
    \\ 
    RoBERTa-base &
    .05 & .65 & .10 & .65 & .08 &
    .04 & .32 & .07 & .32 & .16 &
    .04 & .38 & .08 & .38 & .25 &
    .04 & .55 & .08 & .55 & .18 &
    .04 & .43 & .07 & .43 & .56 &
    .05 & .59 & .10 & .59 & .17
    \\ 
    RoBERTa-large &
    .05 & .61 & .09 & .61 & .10 &
    .05 & .40 & .09 & .40 & .12 &
    .05 & .44 & .09 & .44 & .17 &
    .05 & .66 & .09 & .66 & .12 &
    .04 & .47 & .08 & .47 & .51 &
    .06 & .63 & .10 & .63 & .15
    \\ 
    \midrule
    Llama2-7B &
    .01 & .11 & .02 & .11 & .39 &
    .00 & .02 & .00 & .02 & .37 &
    .00 & .04 & .01 & .04 & .46 &
    .01 & .10 & .01 & .10 & .42 &
    .00 & .03 & .01 & .03 & .96 &
    .01 & .10 & .02 & .10 & .51
    \\ 
    Llama2-13B &
    .00 & .05 & .01 & .05 & .43 &
    .01 & .05 & .01 & .05 & .32 &
    .01 & .06 & .01 & .06 & .50 &
    .01 & .14 & .02 & .14 & .40 &
    .01 & .07 & .01 & .07 & .93 &
    .02 & .23 & .04 & .23 & .42 
    \\ 
    Llama2-70B &
    .00 & .04 & .01 & .04 & .39 &
    .00 & .02 & .00 & .02 & .38 &
    .00 & .03 & .01 & .03 & .47 &
    .00 & .04 & .01 & .04 & .45 &
    .00 & .02 & .00 & .02 & .98 &
    .00 & .05 & .01 & .05 & .53
    \\         
    Mistral-7B &
    .01 & .13 & .02 & .13 & .36 &
    .00 & .03 & .01 & .03 & .31 &
    .01 & .06 & .01 & .06 & .46 &
    .01 & .11 & .02 & .11 & .41 &
    .00 & .04 & .01 & .04 & .96 &
    .01 & .12 & .02 & .12 & .47
    \\ 
    Mixtral-8x7B &
    .01 & .16 & .02 & .16 & .31 &
    .00 & .03 & .01 & .03 & .36 &
    .01 & .06 & .01 & .06 & .47 &
    .01 & .08 & .01 & .08 & .44 &
    .00 & .05 & .01 & .05 & .95 &
    .02 & .18 & .03 & .18 & .42 
    \\ 
    \midrule
    T5-base &
    .06 & .73 & .11 & .73 & .05 &
    .05 & .36 & .08 & .36 & .13 &
    .05 & .46 & .09 & .46 & .16 &
    .04 & .58 & .08 & .58 & .12 &
    .04 & .49 & .08 & .49 & .49 &
    .06 & .67 & .11 & .67 & .11
    \\ 
    T5-large &
    .06 & .72 & .11 & .72 & \underline{\textbf{.04}} &
    .05 & .40 & .09 & .40 & \underline{.11} &
    .05 & .47 & .09 & .47 & .18 &
    .05 & .65 & .09 & .65 & .10 &
    \underline{.05} & .52 & \underline{.09} & .52 & .45 &
    .06 & .71 & \underline{\textbf{.12}} & .71 & .10
    \\ 
    T5-xl &
    .06 & .78 & .11 & .78 & \underline{\textbf{.04}} &
    \underline{.06} & \underline{.44} & \underline{.10} & \underline{.44} & \underline{.11} &
    \underline{.06} & \underline{.52} & \underline{.10} & \underline{.52} & \underline{.15} &
    \underline{.06} & \underline{.77} & \underline{.11} & \underline{.77} & \underline{.07} &
    \underline{.05} & \underline{.53} & \underline{.09} & \underline{.53} & \underline{.44} &
    \underline{.07} & \underline{.76} & \underline{\textbf{.12}} & \underline{.75} & .09 
    \\
\bottomrule

\end{tabular}
}
\endgroup

\end{center}
\label{tab:results_prf}

\end{table*}

\subsection{RQ1: BEST Performing Model}\label{sec:rq1}
The choice of the model used to create the embedding highly affects the zero-shot classifier performance. On $OS_A$ where the top performance was obtained, $F_{189}$ ranged between 0.05 and 0.83 and $D_{n}$ ranged from 0.04 to 0.43, as depicted in Table~\ref{tab:results_prf}. 

The best-performing model varied when using different output spaces to classify requirements. Based on $F_{189}$, the best performing models were: MiniLM (on $OS_A$ and $OS_T$), T5-xl on ($OS_B$, $OS_G$, $OS_K$, $OS_L$ and $OS_T$). However, according to the proposed distance metric ($D_n$), \emph{T5-xl} was the best performing model in all output spaces except for $OS_T$ where it was the second with $D_{n} = 0.09$ compared to $D_{n} = 0.08$ for \emph{Multi-MPNet}. We reject $H_{A0}$ as the model has a significant effect on the classifier performance.

According to the results in Table~\ref{tab:results_prf}, we observe two things regarding $F_{1}$. First, the ranking of the models based on $F_{1}$ varied across different output spaces; four models (MiniLM, DistillRoBERTa, T5-large, T5-xl) were ranked as top models on different occasions. Second, $F_{1}$ is the harmonic mean of precision and recall, and $F_{1}$ is more sensitive to the lower value of the two metrics. Thus, we see that $F_{1}$ is low overall, as the precision is low. This makes the metric $F_{1}$ not a good candidate to select the top-performing model for the zero-shot requirements classifier.
$F_{189}$ ($F_{\beta}$) was better to identify the top models compared to $F_1$. However, two models were ranked as top MiniLM and T5-xl, also $F_{\beta}$ is almost identical to recall, causing precision to be unpresented in $F_{\beta}$. Thus, comparing the results of $F_{\beta}$ is similar to comparing the results of recall in this hairy industrial task. Hence, $F_{\beta}$, where $\beta$ is calculated based on Berry proposal~\cite{berry_empirical_2021}, is in this experiment not useful to identify the top-performing model.

However, as depicted in Table~\ref{tab:results_prf}, the results of the \emph{label distance metric} ($D_n$) were more consistent across different output spaces compared to $F_1$ and $F_{\beta}$. T5-xl was ranked first according to the distance metric across all output spaces, except for $OS_T$, where it was second with 0.01 difference. In conclusion, the \emph{label distance metric} was, in the context of our experiment, a better metric than $F_{1}$ and $F_{\beta}$ to identify the top-performing language model for the zero-shot classifier.

\subsection{RQ2: Model Characteristics Impact}~\label{sec:rq2}

\begin{table}[bth]
    \caption{Statistical tests results}
    \centering
    
    \resizebox{1\columnwidth}{!}{
        \begin{tabular}{lllll}
            \toprule
            Test & Variable & Metric & p-value & $r_s$ \\
            \midrule    
            \multirow{3}{*}{Spearman's Correlation} & \multirow{3}{*}{Model size} & $F_{1}$ &  $<0.05$ & -.646 \\ 
            & & $F_{\beta}$  & $<0.05$ & -.665 \\ 
            & & $D_{n}$   & $<0.05$ & .491\\ 
            \midrule
            \midrule
            \multirow{3}{*}{Kruskal–Wallis H} & \multirow{3}{*}{Embedding size}  
            & $F_{1}$ & $<0.05$ & -\\ 
            & & $F_{\beta}$ & $<0.05$ & - \\ 
            & &  $D_{n}$  & $<0.05$ & -\\ 
            \midrule        
            \multirow{3}{*}{Kruskal–Wallis H} & \multirow{3}{*}{Model type}
            & $F_{1}$ & $<0.05$ & -\\ 
            & & $F_{\beta}$ & $<0.05$ & -\\ 
            & & $D_{n}$ & $<0.05$ & -\\ 
            \bottomrule
        \end{tabular}
    }
    \label{tab:statistics}
\end{table}

We studied the effect of three model characteristics (embedding size, model size, and model type) on  the zero-shot classifier's performance. We performed statistical tests on three performance metrics $F_{1}$, $F_{\beta}$, and $D_{n}$, and present the statistical test results in Table~\ref{tab:statistics}.

\paragraph{Embedding size} 
A Kruskal-Wallis H test was conducted to determine if there were differences in classifier’s performance between the different Embedding sizes: 384 (n=1), 768 (n=7), 1024 (n=1) and 4096 (n=5).  All metrics scores ($F_{1}$, $F_{\beta}$ and $D_{n}$) were statistically significantly different between the different Embedding sizes,  $p<0.05$. The post hoc analysis revealed statistically significant differences in classifier’s performance between embedding size 4096 and all the other embedding sizes ($p < 0.05$), and no difference between the embedding sizes 384, 768 and 1024 ($p > 0.05$). The results are the same for all three metrics ($F_{1}$, $F_{\beta}$ and $D_{n}$). Therefore, we reject $H_{B0}$ as the embedding size significantly impacts the classifier's performance.

\paragraph{Model size} The results from Spearman correlation show that there is a statistically significant correlation between Model size and the classifier’s performance. In particular, the results show that a strong negative correlation exists between Model size and $F_{1}$-score ($r_s$= -.646, $p <0.05$) and between Model size and $F_\beta$ ($r_s$= -.665, $p < 0.05$). On the other hand, a strong positive correlation exists between Model size and $D_n$ ($r_s$= .491, $p <0.05$). Therefore, we reject $H_{C0}$ as correlation exists between the model size and the classifier’s performance. 

\paragraph{Model type} A Kruskal-Wallis H test was conducted to determine if there were differences in classifier’s performance between the different model types: Autorencoding (n=6), autoregression (n=5) and sequence-to-sequence (n=3).  All metrics scores ($F_{1}$, $F_{\beta}$ and $D_{n}$) were statistically significantly different between the different model types,  $p<0.05$. The post hoc analysis revealed statistically significant differences in classifier’s performance between the autoregression and autoencoding model type ($p < 0.05$), and between autoregression and sequence-to-sequence model type ($p < 0.05$), but not between Autorencoding and sequence-to-sequence model type ($p > 0.05$). The results are the same for all three metrics ($F_{1}$, $F_{\beta}$ and $D_{n}$). Consequently, we reject $H_{D0}$ as the model type significantly impacts the classifier's performance.

\section{Discussion}\label{sec:discussion}
\subsection{LM vs LLMs for Semantic Similarity}\label{sec:lm vs llms}
The best models for the zero-shot classifier are the sequence-to-sequence models and autoencoding models. These models were able to detect semantic similarities between requirements text and classes in the domain taxonomies. What these models have in common is that they are based on SBERT architecture, which is developed to generate sentence embeddings that are comparable using cosine-similarity~\cite{reimers-2019-sentence-bert}. Thus, when selecting a model for semantic similarity tasks on requirements artifacts, we should consider using the SBERT variant of the model if it exists, potentially leading to high performance. Moreover, the SBERT-based models are efficient on resources as they could process over 2000 sentences per second on a single GPU~\cite{reimers-2019-sentence-bert}, we run each experiment instance in less than 1 minute on a single GPU with 11 GB of RAM. Consequently, the SBERT-based model are more appealing to adapt for industry with short return on investment (ROI).

Although LLMs showed their superiority over traditional ML and deep learning methods to perform multiple tasks in software engineering~\cite{chang_survey_2024}, they fall short on embedding generation to detect semantic similarity between two texts, as we saw in our results (Table~\ref{tab:results_prf}). A similar observation was made by Tao et al.~\cite{tao_eveval_2023} and Riccardi et al.~\cite{riccardi_two_2023}, where LLMs did not perform well in understanding the semantics of texts and detecting similar ones. However, this can't be solely attributed to the model's size, i.e., the number of parameters that a model learns during pre-training, but also to the type of such models based on the architecture. T5-xl, a sequence-to-sequence model, had the highest performance among all models on the classification tasks despite having a relatively moderate number of parameters (3B) compared to other models that we studied, e.g., Llama2 has the highest number of parameters (70B), but the model performance was one of the lowest, something that we can attribute to being an autoregression model. Autoregression models generate embeddings using a decoder which tokenizes and generates an embedding for each word separately~\cite{radford_improving_2018,wang_what_2022}. It then modifies each word's embedding based on the previous word to generate contextualized embeddings. This is different from how autoencoding and sequence-to-sequence models generate embeddings using an encoder~\cite{kaiser_one_2017}. The encoder tokenizes the text and then looks at both the previous and subsequent tokens to generate an embedding based on the entire text, rather than just depending on the previous token. 

Previous studies~\cite{chang_survey_2024}, evaluating LLMs on tasks from various domains, concluded that LLMs are unable to perform well on all tasks, which aligns with our results. However, we should not conclude that the large number of parameters in LLMs is useless; rather, we should consider the model type when making such conclusions. So, when selecting an LLM for an NLP task, we should consider other model characteristics besides the size, such as model type and the task for which the model is pre-trained or fine-tuned.

\subsection{Model Characteristics}

As we saw in our answer to RQ2 (Section~\ref{sec:rq2}), the statistical tests suggest that the three model characteristics that we studied (embedding size, model size, and model type) have an effect on the classifier performance. 

The negative correlation between the embedding size and the classifier’s performance can be attributed to the large size of embedding generated by autoregression models. As we discussed in Section~\ref{sec:lm vs llms}, these autoregression models have the worst performance compared to other types of models. Due to that, and as the correlation was weak, we did not make a conclusion about the effect of embedding size and the classifier’s performance. 

We make a similar observation regarding model size, where it negatively correlates with the classifier performance, although models with larger sizes are expected to perform better~\cite{chang_survey_2024}, as they are trained on larger datasets and learn more parameters. As discussed in Section~\ref{sec:lm vs llms}, these LLMs are normally pre-trained and fine-tuned for language generation rather than embedding generation. Furthermore, the label distance metric, as depicted in Figure~\ref{sec:distance}, does not show a clear correlation with the parameters count. Thus, the model type has a stronger effect than the model size on the performance of the zero-shot classifier when the models are used to generate embedding for semantic similarity.  

\subsection{Label Distance vs. Traditional Metrics}
As mentioned in Section~\ref{sec:rq1} and based on the results in Table~\ref{tab:results_prf}, the top-performing model could not be easily identified using $F_1$ or $F_{\beta}$. However, comparing the results of the proposed metric \emph{$D_n$}, T5-xl is the top-performing model on all output spaces, except on \emph{$OS_T$}, where it came second with 0.01 difference. Consequently, the results of the distance metric were more consistent across multiple datasets, compared to $F_1$ and $F_{\beta}$. Furthermore, the distance metric is more granular and could identify small differences between models that are unclear with other metrics. 

In a recommender system for hierarchical text classification, the distance metric can be beneficial in identifying the neighboring classes to display to the user in addition to the recommended classes. This improves the probability of finding the true class(s) in the output of the recommender without the need to traverse through the whole hierarchy.

\subsection{Label Distance Metric Usage}
The \emph{label distance metric} is meant to be a more granular metric compared to recall, precision, $F_{1}$-score and $F_{\beta}$ to evaluate the classifier performance and select the top-performing model. A lower distance means better performance and a zero distance means 100\% recall. However, the acceptable distance value could vary depending on the context where the model will be used. For example, in our case, we intend to use the zero-shot classifier to recommend classes from a taxonomy to a requirement text. An expert in the field (e.g., requirement engineer) will sanitize these recommendations to choose the correct labels. When building such a recommender system, during the design phase, the absolute label distance ($D_a$) can be used to determine, on average, how close the predicted labels are to the true labels. Consequently, we could show the predicted labels from the taxonomy with their neighboring classes (i.e., parent, children, and siblings) that have a distance of $D_a$ from a predicted label. Thus, it becomes easier for the expert to select the true label as they would only need to look at the part rather than the whole taxonomy when classifying the requirement text.

\section{Conclusion and Future Work}\label{sec:conclusion}

In this study, we evaluated the applicability of a zero-shot classifier that uses language models on a multi-label requirements classification task. We experimented with various LMs and LLMs to evaluate the classifier’s performance using a dataset annotated with six large hierarchical output spaces. In addition, we presented a novel metric, \emph{label distance}, a more fine-grained metric that is useful for model selection. Our results show that a larger model size would not necessarily lead to higher classifier performance; rather, the model type has a larger effect on the performance, and autoencoding and sequence-to-sequence models have the highest performance. Moreover, with the \textit{label distance metric}, we could identify the best-performing model across multiple datasets. 

The zero-shot classifier is a promising option for performing multi-label requirements classification using large taxonomies in multiple domains. The classifier could be used to semi-automate the requirements classification process in practice, where an expert would vet the classifier's output and select the correct labels.  

In the future, a thorough evaluation of the proposed metric must be conducted to ensure that the metric results are useful for making decisions when building an automated classification tool. Also, we encourage researchers to conduct more studies in the area of multi-label requirements classification, as addressing this problem could pave the way to other RE activities, e.g., requirements smell detection or trace-link recovery.

\section{Data Availability}\label{sec:data availability}
We provide our replication package~\footnote{https://doi.org/10.6084/m9.figshare.25990825.v1} including the source code and the dataset. We have published 342 labeled requirements from the publicly available data. The remaining 35 requirements are project-specific and are not published due to the protection of company data. However, the remaining requirements can be made available upon a reasonable request.

\balance
\bibliographystyle{IEEETrans}
\bibliography{IEEEabrv,refrences}

\end{document}